\documentclass[twocolumn]{aastex62}
\usepackage{graphicx}

\begin{document}
\title{\bf Kinematic evidence for an embedded protoplanet in a circumstellar disc}

\author[0000-0001-5907-5179]{C. Pinte}
\email{christophe.pinte@monash.edu}
\affiliation{Monash Centre for Astrophysics (MoCA) and School of Physics and Astronomy,
  Monash University, Clayton Vic 3800, Australia}
\affiliation{Univ. Grenoble Alpes, CNRS, IPAG, F-38000 Grenoble, France}
\author[0000-0002-4716-4235]{D.\,J.~Price}
\affiliation{Monash Centre for Astrophysics (MoCA) and School of Physics and
  Astronomy, Monash University, Clayton Vic 3800, Australia}
  \author[0000-0002-1637-7393]{F.~M\'enard}
\affiliation{Univ. Grenoble Alpes, CNRS, IPAG, F-38000 Grenoble, France}
\author[0000-0002-5092-6464]{G.~Duch\^ene}
\affiliation{Astronomy Department, University of California, Berkeley,
  CA 94720-3411, USA}
\affiliation{Univ. Grenoble Alpes, CNRS, IPAG, F-38000 Grenoble, France}
\author{W.\,R.\,F.~Dent} \author[0000-0003-3253-1255]{T. Hill} \author[0000-0003-4518-407X]{I. de Gregorio-Monsalvo}
\affiliation{Atacama Large Millimeter / Submillimeter Array, Joint ALMA
  Observatory, Alonso de C\'ordova 3107, Vitacura 763-0355, Santiago, Chile}
\author[0000-0001-5073-2849]{A. Hales}
\affiliation{Atacama Large Millimeter / Submillimeter Array, Joint ALMA
  Observatory, Alonso de C\'ordova 3107, Vitacura 763-0355, Santiago, Chile}
\affiliation{National Radio Astronomy Observatory, 520 Edgemont Road, Charlottesville, VA 22903-2475, United States of America}
\author[0000-0002-5526-8798]{D. Mentiplay}
\affiliation{Monash Centre for Astrophysics (MoCA) and School of Physics and Astronomy,
  Monash University, Clayton Vic 3800, Australia}

\begin{abstract}
 Discs of gas and dust surrounding young stars are the birthplace of
 planets. However, the direct detection of protoplanets forming within discs has
 proved elusive to date. We present the detection of a large, localized
 deviation from Keplerian velocity in the protoplanetary disc surrounding the
 young star HD~163296. The observed velocity pattern is consistent with the
 dynamical effect of a two-Jupiter-mass planet orbiting at a radius $\approx$ 260\,au from the star.
\end{abstract}

\keywords{stars: individual (HD~163296) --- protoplanetary discs  ---
  planet-disc interaction --- submillimeter: planetary systems --- hydrodynamics --- radiative transfer}

\section{Introduction}

\begin{figure*}
  \centering
  \includegraphics[width=\hsize]{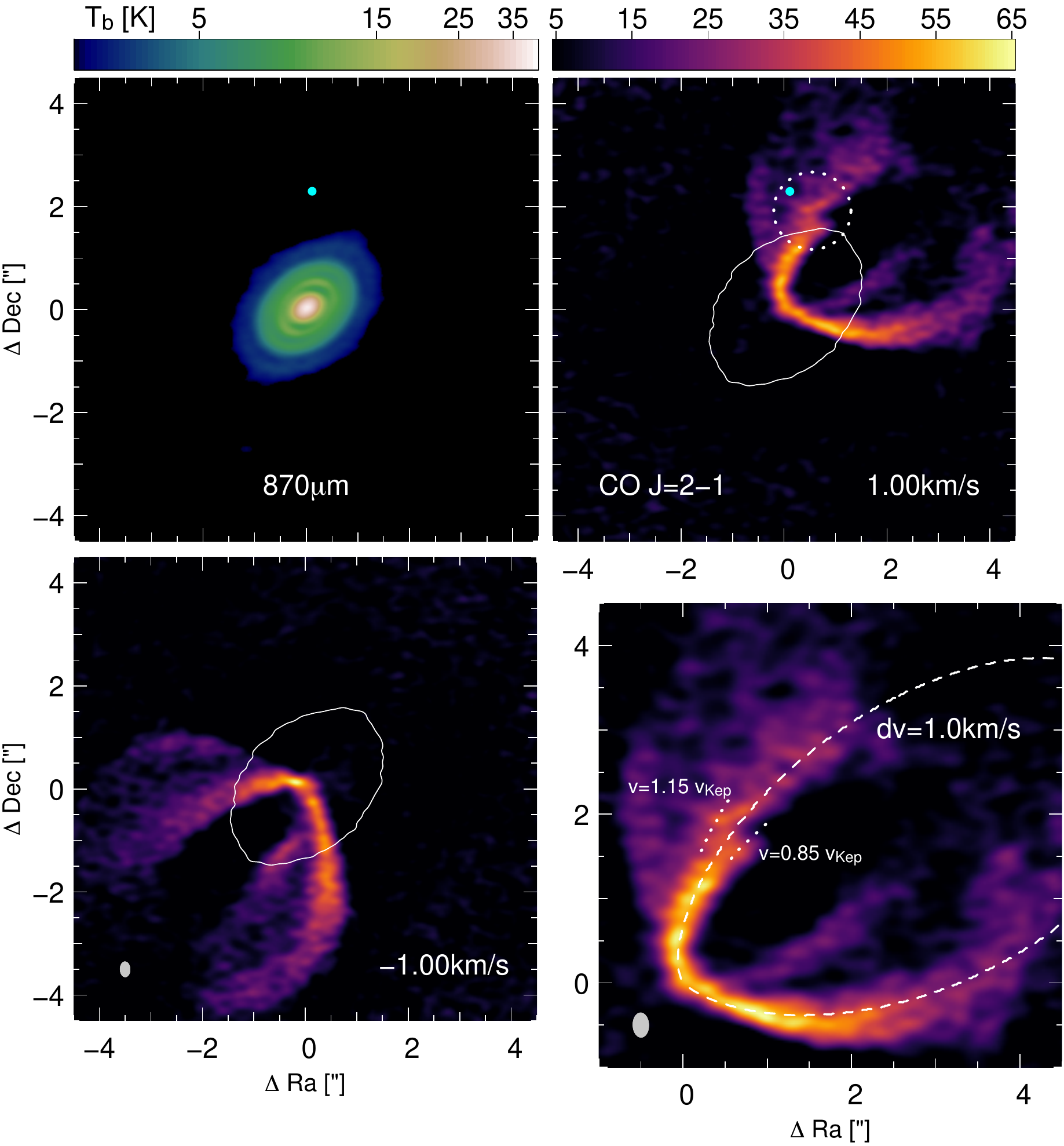}
  \caption{Kinematic asymmetry in HD~163296. Band 6 continuum emission
    (top left) and
    channel map of $^{12}$CO line emission at $+1$km s$^{-1}$ from the systemic
    velocity (top right, with a close-up shown in bottom right) shows a distinct
    `kink' in the emission (highlighted by the dotted circle). Comparison with the
    continuum emission (top left) locates this outside of the outermost dust
    ring. The corresponding emission on the opposite side of the disc (bottom
    left; showing $-1$km s$^{-1}$ channel) shows no corresponding feature, indicating
    the disturbance to the flow is localised in both radius and
    azimuth. The channel width is  $\Delta v = 0.1$\,km s$^{-1}$. The
      white contour shows the 5-$\sigma$ ($\sigma= 0.1$\,mJy beam$^{-1}$) level of the continuum map. The dashed
      line is the expected location of the isovelocity curve on the upper
      surface of a disc with an opening angle of $15^\circ$ and an inclination of
      $45^\circ$. Dotted
    lines in the bottom-right figure indicate 15\% deviations ($\approx$
    0.4\,km s$^{-1}$) from Keplerian
    flow around the star. The potential planet location is marked by a cyan
    dot, assuming it is located in the midplane. \label{fig:one_channel}}
\end{figure*}

Direct observations of forming planets in protoplanetary discs is the ultimate
goal of disc studies. The disc usually outshines the planet, requiring
observations at high contrast and angular resolution. Detections by direct
imaging have been reported in several discs: HD~100546
\citep{Quanz2013a,Brittain2014,Quanz2015,Currie2015}, LkCa~15
\citep{Kraus2012,Sallum2015}, HD~169142
\citep{Quanz2013b,Biller2014,Reggiani2014}, and MWC 758
\citep{Reggiani2018}. Yet, most of the detections to date have been
subsequently challenged \citep[e.g.,][]{Thalmann2015, Thalmann2016, Rameau2017,
  Ligi2018}. The quest continues.

An alternative approach is to search for indirect signatures imprinted by
planets on their host disc. The Atacama Large Millimetre Array (ALMA), and adaptive optics systems have revealed a
variety of structures: gaps and rings \citep{ALMA-Partnership2015,Andrews2016,Isella2016a}, spirals \citep[e.g.][]{Benisty2015,Stolker2016}, that could be signposts of
planets, but numerous other explanations also exist that do not require planets \citep[e.g.][]{Takahashi2014,Flock2015,Gonzalez2015,Loren2015,Zhang2015,Bethune2016}.
Embedded planets in circumstellar discs will launch spiral waves at
Lindblad resonances both inside and outside of their orbit
\citep[e.g.][]{Ogilvie2002}, disturbing the local Keplerian velocity
pattern. Hydrodynamic simulations show that the impact
on the velocity pattern should be detectable by high spectral resolution ALMA line observations
\citep{Perez2015b}.  Deviations from Keplerian rotation have been
detected around circumbinary discs, with streamers at near free-fall
velocities \citep{Casassus2015a, Price2018} and radial flows or warps \citep{Walsh2017}.


HD 163296 is a $\sim$4.4Myr old Herbig Ae star located at a distance of $101.5
\pm 1.2$\,pc from the Sun \citep{Gaia-Collaboration2018}. We rescaled all relevant quantities from previous papers based on the new Gaia distance.
HD~163296 has a mass of $1.9$\,M$_{\odot}$ \citep[e.g.][]{Flaherty2015}, a
luminosity of $25$\,L$_{\odot}$ \citep{Natta2004},
  and an A1Ve
spectral type, with effective temperature 9300\,K.
Observations with the Hubble Space Telescope (HST) revealed a disc in scattered light that extends as far out as 375\,au
\citep{Grady2000}. Interestingly, \citet{Grady2000} inferred the presence of a
giant planet at $\approx$ 270\,au based on the gap observed in scattered light at that radius.
\citet{de-Gregorio-Monsalvo2013} presented ALMA data and showed that the
gaseous component of the disc extends to distances of at least
R$_\mathrm{out-CO}=$415\,au in CO while the
continuum is detected only to
R$_\mathrm{out-Dust}=$200\,au. Higher
resolution ALMA imaging revealed a bright inner disc component within the inner
$0.\! ''5$, and a spectacular series of three rings at
$\approx$ 65\,au, 100\,au, with a fainter ring at
160\,au \citep{Isella2016a}.

\begin{figure*}
  \includegraphics[width=\hsize]{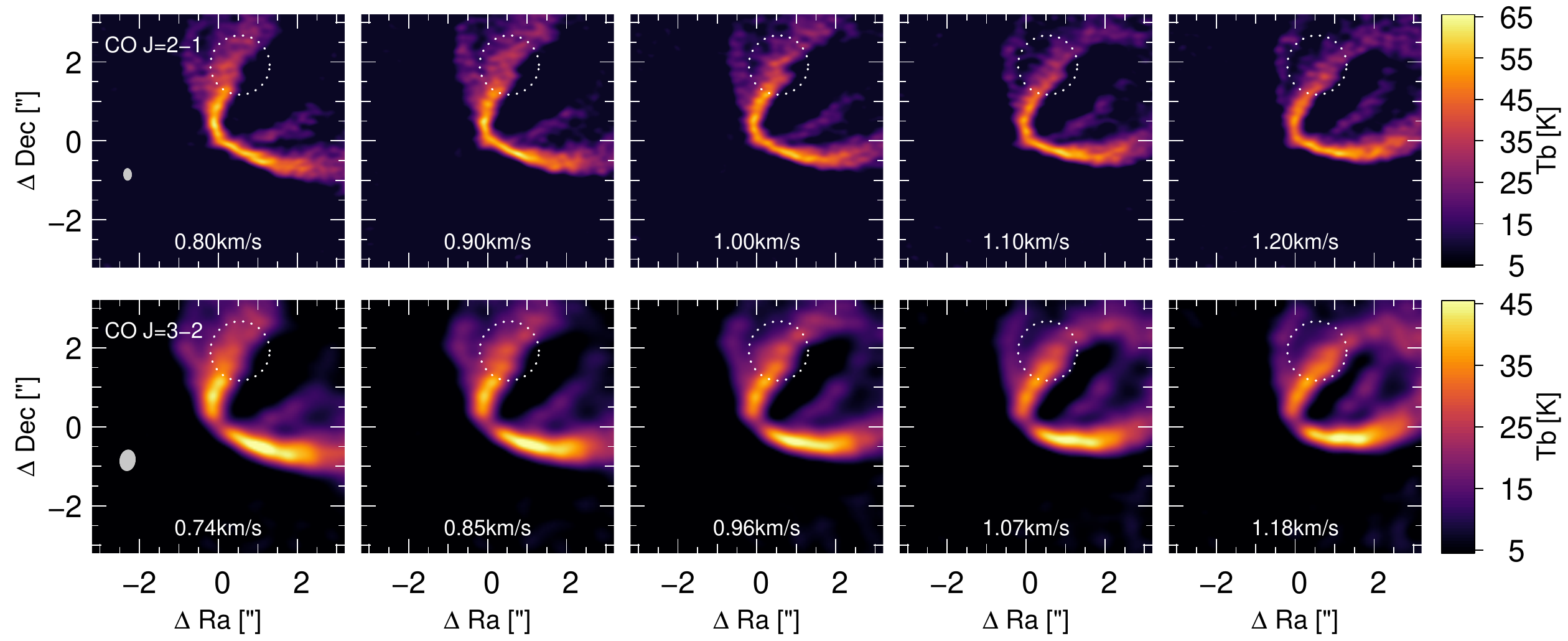}
  \caption{Channel maps around the detected deviation from Keplerian
    velocity. The `kink' is most visible in channels at velocities between 0.8
    and 1.2 km/s (top row) and is also seen in the J=3--2 transition in similar velocity channels (bottom row) indicating it is localised in both space and velocity.\label{fig:channel_maps}}
\end{figure*}

In this Letter, we present the detection of a local deviation from the
Keplerian velocity pattern found in high spectral resolution ALMA imaging. By
comparing with models we find this to be consistent with the presence
of a few-Jupiter mass protoplanet in the disc.

\section{Observations and data reduction}

We use archival ALMA data. Observations were performed on 2012 June 9, 11, 22,
and July 6 at Band 7 (2011.0.000010.SV), and on 2015 August 5, 8, and 9 at Band
6 (2013.1.00601.S). A complete description of the data was presented in
\cite{de-Gregorio-Monsalvo2013} and \cite{Isella2016a}. For the Band 7 data, we
re-used the maps produced by \cite{de-Gregorio-Monsalvo2013}, with a $0.52''
\times 0.38''$ beam at PA=82$^\circ$, and a channel width of
0.11\,km\,s$^{-1}$.

We used {\sc casa} scripts provided by ALMA to calibrate the Band 6 data. As
the data from the night of August 9 showed significantly
higher noise and flux levels, we selected only the data from the August 5 and 8
for the analysis. We performed three successive
rounds of phase self-calibration, the last with solutions calculated for each
individual integration (6s), followed by a phase and amplitude
self-calibration. The continuum self-calibration solutions were applied to the
CO lines.
Imaging was performed at 0.1 km s$^{-1}$ resolution, using Briggs weighting
with a robust parameter of -0.5 to obtain a synthesized beam of $0.28'' \times
0.18''$ at PA=-88$^\circ$. We did not subtract the continuum emission in order to avoid
underestimating the gas temperature and affecting the apparent
  morphology of the emission \citep[e.g.][]{Weaver2018}. At the location of the
  detected velocity deviation, continuum emission is negligible, and an
  analysis on continuum-subtracted data would lead to the same results.

\section{Results and analysis}

\begin{figure}
  \centering
  \includegraphics[width=0.7\hsize]{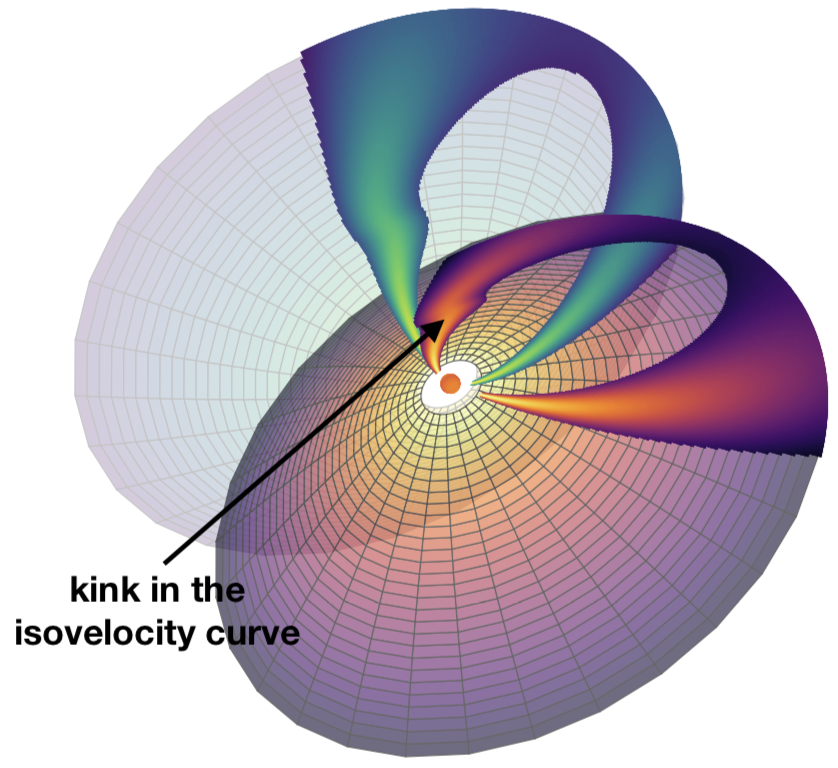}
\caption{Geometry of the inclined and flared disc, showing a schematic of the expected emission from two infinitely thin emitting surfaces.
  Green shows the emission from the lower surface of the disc, and red shows the upper surface. We added a 10\%
  deviation in azimuthal velocity north of the star, which appears as a `kink' in the line emission. Emission is only seen when the projected velocity matches the channel velocity, producing the characteristic `butterfly' shape. Emission is preferentially seen on the upper surface of the disc due to the higher inclination with respect to the line of sight.
\label{fig:toy_model}}
\end{figure}

The disc shows the typical butterfly pattern of discs in Keplerian rotation
\citep{de-Gregorio-Monsalvo2013, Rosenfeld2013a}. In a given channel, the
emission is concentrated along an isovelocity curve, corresponding to the
region of the disc where the projected velocity is equal to the channel
velocity.
The emission from the upper and lower surfaces --- above and below the midplane
as seen by the observer --- and from the near and far sides of these surfaces,
is well separated (Fig.~\ref{fig:one_channel}, and schematic view in Fig.~\ref{fig:toy_model}).

In a recent paper \citep{Pinte2018} we showed how to reconstruct the position
and velocity of each of the CO layers, for discs at intermediate inclination,
by simple geometrical arguments based on
the emission in each channel map. HD~163296 displays a similar scale height
and velocity profile to the T Tauri star IM Lupi \citep{Pinte2018}, with a
flared CO emitting surface and decreasing velocities and temperature with
radius (C. Pinte et al., in prep.).

Significantly, HD~163296 shows
an asymmetry between
  the southeast and northwest sides of the disc
at a cylindrical radius of $\approx$ 260\,au, outside of the
third dust ring seen in continuum emission.
This asymmetry is most evident in channels at a projected velocity of
$\approx 6.8 \pm 0.2$\,km s$^{-1}$ ($\approx 1$\,km s$^{-1}$ from the systemic velocity).
Fig.~\ref{fig:one_channel} shows the corresponding individual velocity
channels. The emission feature --- highlighted by the dotted circle ---
corresponds to a kink in the upper-surface isovelocity curve northwest of the
central object at velocities close to ${\rm d}v=+1\,$\,km s$^{-1}$. The
symmetric channel (${\rm d}v=-1\,$\,km s$^{-1}$,) shows a smooth Keplerian
profile to the southeast.
We detect a similar deformation of the isovelocity curves at the same location
in both $^{12}$CO $J=2-1$ and $3-2$ transitions
(Fig.~\ref{fig:channel_maps}). While it is not as obvious in the Band 7
Early-Science data due to the limited spatial resolution,  the deformation of
the isovelocity curve is present and could already be seen, with the benefit of
hindsight from our Band 6 detection, in \cite{de-Gregorio-Monsalvo2013} and
\citet[][their Fig.~3 and 2, respectively]{Rosenfeld2013a}. The asymmetry is not detectable in the less
abundant isotologues $^{13}$CO and C$^{18}$O, where the emission is more
diffuse and fainter because of the lower optical depth.

The deformation of the emission is localised to an area approximately $0.5''$
in size (indicated by the dotted circle in Figures~\ref{fig:one_channel}
and~\ref{fig:channel_maps}) and to channel maps at velocities between 0.8 and
1.2\,km s$^{-1}$ from the systemic velocity (top row of
Fig.~\ref{fig:channel_maps}). This argues for a localised perturbation and
excludes an origin from any large-scale structure in the disc.

\section{Models and discussion}

The detected asymmetry matches our expectations for a local deviation from
Keplerian velocity caused by a massive body embedded in the disc. A local
deviation of $\approx$ 0.4\,km s$^{-1}$ is enough to reproduce the observed
spatial shift
(Fig.~\ref{fig:one_channel}, bottom-right panel). The dotted lines shown in the bottom-right panel delineate what would be $\approx$ 15\% deviations in the local velocity field, which is the approximate extent of the deviation from Keplerian rotation. Most significantly, the shape of the deviation in the emission maps is similar to the prediction by \cite{Perez2015b} for the kinematic signatures of an embedded planet, where the wake of the spiral generated by the planet was shown to produce a kink in the emission due to the deviation from the Keplerian rotation around the central star.

The basic feature of the channel maps can be explained with a simple model
assuming emission from two infinitely thin emitting
surfaces. Figure~\ref{fig:toy_model} shows the expected emission arising from
such a model, showing the butterfly signature from the disc. Asymmetries of the
velocity field, added in an ad hoc manner in the model for illustrative purposes, are evident as small bumps on the line emissions.

To go beyond this simple model and infer the mass of the putative planet, we
performed a series of 3D global simulations using the {\sc phantom} Smoothed
Particle Hydrodynamics (SPH) code \citep{Price2017}. We adopted the gas disc
parameters from \cite{de-Gregorio-Monsalvo2013}.
We employed gas-only simulations, ignoring the
effect of dust, using 1 million SPH particles and a central mass of
1.9\,M$_\odot$. The inner radius of the disc in our model was set to 50\,au
(mainly to speed up the calculations as the inner disc is irrelevant for our
present purpose), with an initial outer radius set to 500\,au. We set the gas
mass between those radii to $10^{-2}$\,M$_\odot$, and use an exponentially
tapered power-law surface density profile with a critical radius of 100\,au,
power-law index of $p = -1.0$, and an exponent $\gamma =$ 0.8. The disc aspect
ratio was set to 0.08 at 50\,au, with a vertically isothermal profile.
We set the artificial viscosity in the code in order to obtain an average
\cite{ShakuraSunyaev1973} viscosity of $10^{-3}$ \citep{Lodato2010}, in
agreement with the upper limits found by \cite{Flaherty2015,Flaherty2017}.

\begin{figure*}
  \begin{center}
  \hfill
  \includegraphics[width=0.45\textwidth]{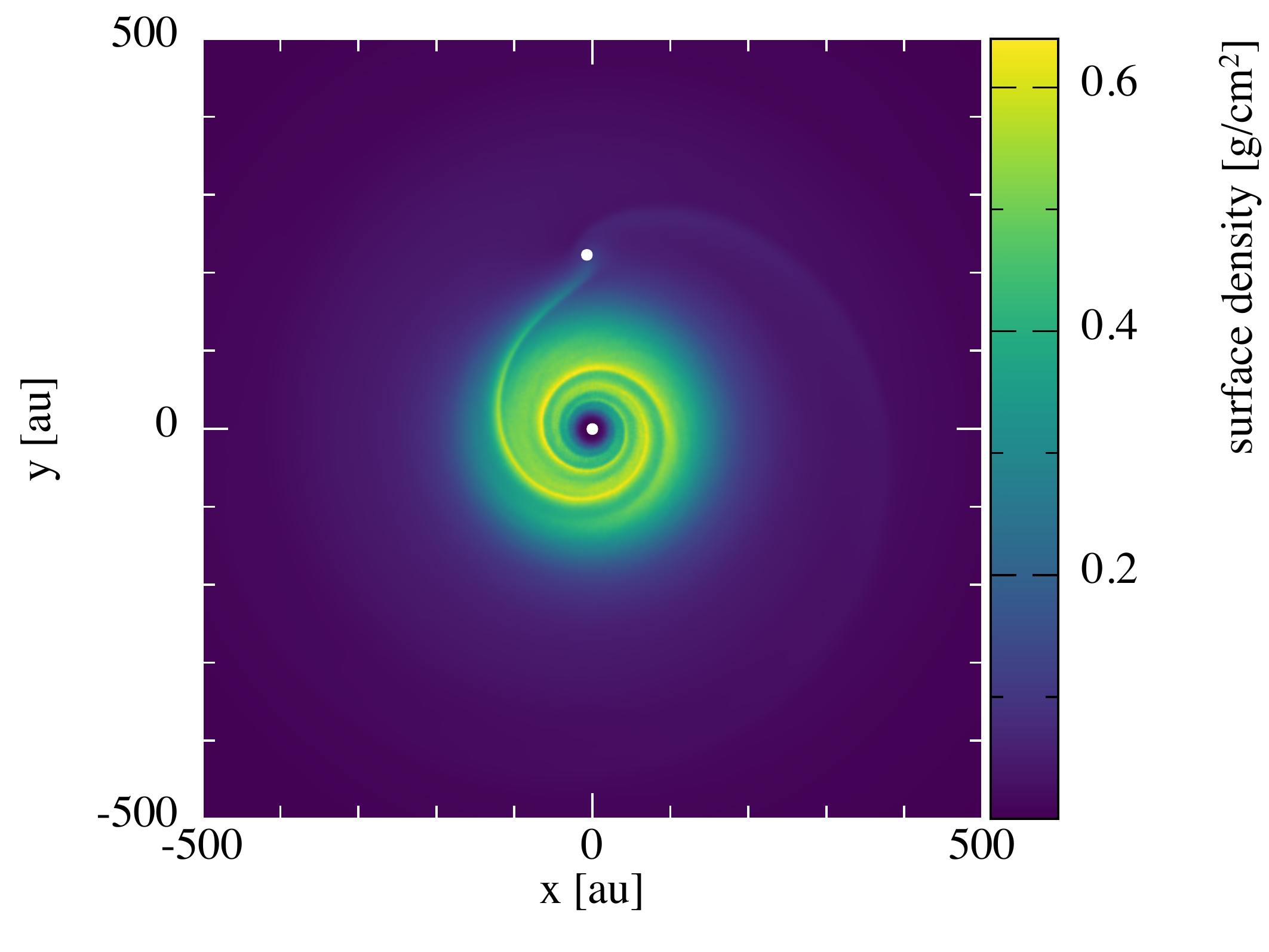}
  \hfill
  \includegraphics[width=0.45\textwidth]{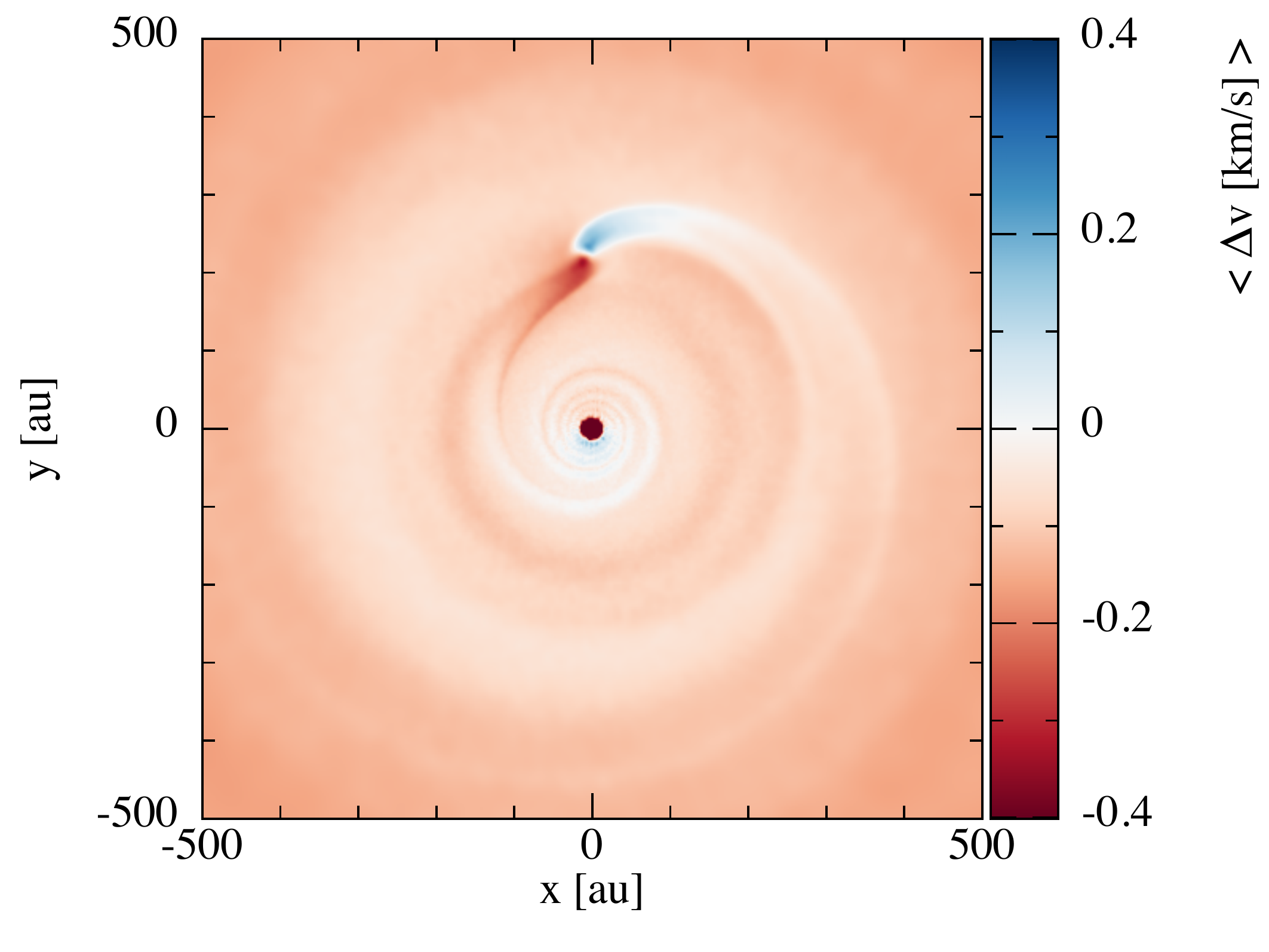}
  \hfill
\caption{Left panel: surface density in 3D hydrodynamics simulations of the
  HD~163296 disc, shown after 35 orbits of a $2$\,M$_\mathrm{jup}$ planet
  and viewed at a face-on inclination. Dots mark the star and
  planet. Right panel: deviation of the azimuthal velocity from Keplerian velocity.\label{fig:phantom}}
\end{center}
\end{figure*}

We embedded a single planet in the disc orbiting at 260\,au with a mass of
either 1, 2, 3, or 5\,M$_{\rm Jup}$. We used sink particles \citep{Bate95} to
represent the star and planet. We set the accretion radius of the planet to
half of the Hill radius (7.05, 8.85, 10.15 and 12\,au, respectively), with an
accretion radius of 10\,au for the central star. The model surface
density is plotted in Fig.~\ref{fig:phantom} for the 2\,M$_\mathrm{jup}$
planet. We evolved the models for 35 orbits of
the planet ($\approx 100,000$ years), which is sufficient to establish
the flow pattern around the planet.

\begin{figure*}
\includegraphics[width=\hsize]{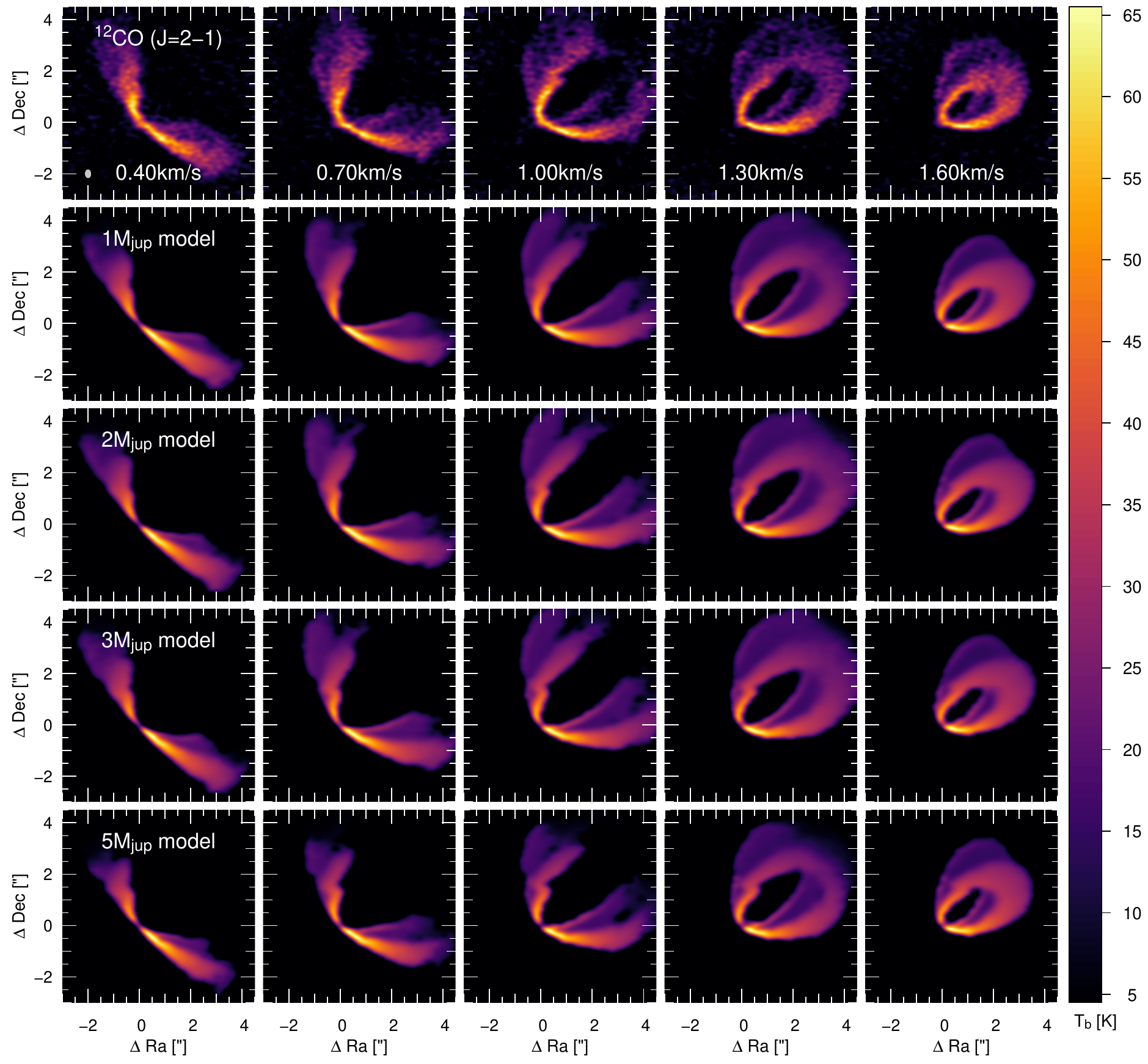}
\caption{Comparison of $^{12}$CO J=2--1 ALMA observations (top row) with synthetic channel maps
    from our 3D hydrodynamics calculations with
  embedded planets of 1, 2, 3 and 5 M$_\mathrm{Jup}$ (from top to bottom). Channel
  width is 0.1km s$^{-1}$. Synthetic maps were convolved to a Gaussian beam to
  match the spatial resolution of the observations
\label{fig:mcfost-phantom}.}
\end{figure*}

To compute the temperature and synthetic line maps, we used the {\sc mcfost} Monte Carlo radiative
transfer code \citep{Pinte06,Pinte09}, assuming
$T_\mathrm{gas} = T_\mathrm{dust}$, and local thermodynamic equilibrium as we are looking at low-$J$ CO lines. The central star was represented by a sphere
of radius 2.1\,R$_\odot$, radiating isotropically with a Kurucz
spectrum at 9,250\,K. We used a Voronoi tesselation where each cell corresponds
to an SPH particle, avoiding the need to interpolate between the SPH and
radiative transfer codes. We set the CO abundance following the prescription in
Appendix B of \cite{Pinte2018} to account for freeze-out where T $< 20$\,K, as
well as
photo-dissociation and photo-desorption in locations where the UV radiation is
high. We adopted a turbulent velocity of 50 m\,s$^{-1}$, consistent with the upper limits
found by \cite{Flaherty2015} and \cite{Flaherty2017}. We assumed a population
of astrosilicate \citep{Draine03} grains with sizes ranging from 0.03 to
1000$\mu$m and following a power-law
$\mathrm{d}n(a) \propto a^{-3.5}\mathrm{d}a$,  a gas-to-dust ratio of 100, and
computed the dust optical properties using Mie theory.

Figure~\ref{fig:mcfost-phantom} presents the predicted emission in $^{12}$CO $J=2-1$ of our theoretical models for
four different planet masses. A 2\,M$_\mathrm{jup}$
planet appears to reproduce a deformation of the $^{12}$CO
isovelocity curve that is consistent with the observations. At
1\,M$_\mathrm{jup}$, the planet only produces a small deformation that is barely
visible in the channel maps, while a more massive planet triggers a strong
spiral arm that would have been detected in channel maps at least up to
0.5\,km s$^{-1}$ from the nominal velocity of 1\,km s$^{-1}$.
The twisted emission in the channel maps is a direct consequence of
  deviation from Keplerian velocity generated by the planet along the wake of
  the spiral arms (Fig.~\ref{fig:phantom}, right panel). \cite{Perez2015b} also predicted
  that the circumplanetary disc can be detected as a compact emission separated
  in velocity from the circumstellar disc emission. The circumplanetary disc
  radius is about one-third of the Hill radius \citep[e.g.][]{Ayliffe2009}. A
  2--5 Jupiter mass planet would produce a circumplanetary disc with a diameter
  smaller than 6--8\,au, respectively. At the current spatial
  resolution of the ALMA observations, its flux will be diluted in the beam ($\approx$ 20\,au).

Note that for the adopted disc parameters, the planet
  migrates by about 30\,au during the simulation, and the synthetic maps
  display the velocity deviation slightly closer to the star than in the data. At this
  rate, the planet would reach the star in about 1\,Myr (though we overestimate
  the migration rate by a factor of 2--3 due to the relatively large sink
  particle radius that we adopted; see \citealt{Ayliffe2010}). If the detection is confirmed, the
  survival of such an  embedded planet could put additional constraints on the
  disc surface density profile and viscosity.

\cite{Grady2000} detected a gap in the scattered light images with
  HST/STIS at 260\,au, and estimated the mass of a
  potential planet to be 0.4\,M$_\mathrm{jup}$ (based on some simple analytical
  derivation). \cite{Isella2016a} also detected a small dip in the integrated CO
  brightness profile at $\approx 2.2$'' (see their Figure~1, or Figure~5 in
  \citealp{Liu2018}). In our model, the gap appears in scattered light for a planet
  mass larger than 2\,M$_\mathrm{jup}$, but remains undetected in the synthetic
  CO maps. The final profile of a planetary gap establishes itself on a viscous
  timescale however (thousands of orbits with a viscosity of $10^{-3}$),
  however, and
  the gap width and depth in our models are only lower limits.

The effect of the planet appears fainter in the  $^{13}$CO channels maps than in
the $^{12}$CO maps, even if the planet is located in the
midplane. This is due to optical depth and vertical temperature gradient
effects: the $^{12}$CO is coming from a vertically narrow and warm layer
above the midplane, while the $^{13}$CO is originating from a deeper, thicker
layer, where the disc is almost vertically isothermal, resulting in a
uniform emissivity that washes out some of the kinematics signal.

Are we seeing the signature of an embedded planet?
 Can we exclude wishful thinking? The strongest evidence is that the
 perturbation to the disc kinematics is highly localised in both space and
 velocity. This excludes any mechanism that merely produces axisymmetric rings
 in discs. This excludes, for example, ice lines \citep{Zhang2015},
 self-induced dust traps \citep{Gonzalez2015}, instabilities
 \citep{Takahashi2014,Loren2015} and zonal flows \citep{Flock2015}. A spiral
 wave could in principle result from the disc self-gravity, but multiple,
 large-scale  spirals would be expected in that case \citep[e.g.][]{Dipierro2015a} which the
 localized deviation seen in HD~163296 would seem to exclude.

 The localised
 nature of the kinematic perturbation, that it occurs close to the gap found by
 \cite{Grady2000}, and the
 similarity to the signatures predicted by our hydrodynamic
   models is
   a strong evidence for a young protoplanet in
 a gas-rich disc. However, confirmation with direct imaging is the only way to
 be sure. The relatively large planet mass and its known location in the disc
 means direct imaging follow-up might be able to detect it, depending on how
 embedded it is in disc. So far, no point source has so far been detected at the
   location of the potential planet with near-IR adaptive optics systems.
 A 2\,M$_\mathrm{Jup}$ planet is consistent with the upper limits (for an
   unobscured planet) obtained by
 adaptive optics systems, such as SPHERE \citep{Muro-Arena2018} and Keck L'
\citep{Guidi2018}.
Using the formalism developed in \cite{Pinte2018}, we find that the velocity
kink is located at a distance of $\approx$ 260\,au, and an elevation above the
midplane of $\approx$ 70\,au. Assuming that the potential planet is located in the
midplane, it would be at a projected distance of 2.3$\pm$0.2" and  PA
= -3$\pm$5$^\circ$ from the star, where we estimated the
  uncertainties by locating the velocity deviation with half of a beam
  accuracy. If the planet orbit is slightly inclined compared to the disc's
  plane, the position on the sky will be shifted along a line going from
  the northeast to the southwest directions.

 Can massive planets form at a distance of 250\,au from the star? The
 location of giant planets in the outer regions of discs would be broadly
 consistent with gravitational instability. On the other hand, the timescale for
 core accretion may also be reasonable given that HD~163296 is a relatively old
 disc ($\approx$ 5\,Myr). The planet may also have undergone outward
 migration, depending upon the initial profile of the disc.
 It is beyond the scope of this Letter to speculate further.

\section{Summary}

  We detected a 15\% deviation from Keplerian flow around the star in the disc
  around HD~163296. The deviation was detected in both Band~6 and Band~7 in two
  different transitions of $^{12}$CO and matches the kinematic signature
  predicted for an embedded protoplanet.

 Comparing the observations to a series of 3D hydrodynamic and radiative
  transfer models of embedded planets suggests that the kinematic feature is caused
  by a planet of  of $\approx$ 2 M$_{\rm Jup}$ in the midplane. Such a planet
  would be located at a distance of $\approx$ 260\,au.

 \section*{Acknowledgments}
 This Letter makes use of the following ALMA data: ADS/JAO.ALMA\#2011.0.00010.SV
 and ADS/JAO.\-ALMA\#2013.1.00601.S. ALMA is a partnership of ESO (representing
 its member states), NSF (USA) and NINS (Japan), together with NRC (Canada),
 MOST and ASIAA (Taiwan), and KASI (Republic of Korea), in cooperation with the
 Republic of Chile. The Joint ALMA Observatory is operated by ESO, AUI/NRAO and
 NAOJ. The National Radio Astronomy Observatory is a
facility of the National Science Foundation operated under cooperative
agreement by Associated Universities, Inc.
This work was performed on the ozSTAR national facility at Swinburne University
of Technology. ozSTAR is funded by Swinburne and the Australian Government's
Education Investment Fund. We thank the anonymous referee for insightful
comments and suggestions.
C.P. and D.J.P. acknowledge funding from the Australian Research Council via
FT170100040, FT13010003,4 and DP180104235. F.M. and C.P. acknowledge funding from
ANR of France (ANR-16-CE31-0013).

\vspace{5mm}
\facilities{ALMA.}

\software{{\sf CASA} \citep{McMullin07},
         {\sf phantom} \citep{Price2017},
          {\sf splash} \citep{Price2007},
          {\sf mcfost} \citep{Pinte06,Pinte09}.
          }

\bibliographystyle{aasjournal}
\bibliography{HD163296_planet}

\begin{thebibliography}{}
\expandafter\ifx\csname natexlab\endcsname\relax\def\natexlab#1{#1}\fi
\providecommand{\url}[1]{\href{#1}{#1}}
\providecommand{\dodoi}[1]{doi:~\href{http://doi.org/#1}{\nolinkurl{#1}}}
\providecommand{\doeprint}[1]{\href{http://ascl.net/#1}{\nolinkurl{http://ascl.net/#1}}}
\providecommand{\doarXiv}[1]{\href{https://arxiv.org/abs/#1}{\nolinkurl{https://arxiv.org/abs/#1}}}

\bibitem[{{ALMA Partnership} {et~al.}(2015){ALMA Partnership}, {Brogan},
  {P{\'e}rez}, {Hunter}, {Dent}, {Hales}, {Hills}, {Corder}, {Fomalont},
  {Vlahakis}, {Asaki}, {Barkats}, {Hirota}, {Hodge}, {Impellizzeri}, {Kneissl},
  {Liuzzo}, {Lucas}, {Marcelino}, {Matsushita}, {Nakanishi}, {Phillips},
  {Richards}, {Toledo}, {Aladro}, {Broguiere}, {Cortes}, {Cortes}, {Espada},
  {Galarza}, {Garcia-Appadoo}, {Guzman-Ramirez}, {Humphreys}, {Jung}, {Kameno},
  {Laing}, {Leon}, {Marconi}, {Mignano}, {Nikolic}, {Nyman}, {Radiszcz},
  {Remijan}, {Rod{\'o}n}, {Sawada}, {Takahashi}, {Tilanus}, {Vila Vilaro},
  {Watson}, {Wiklind}, {Akiyama}, {Chapillon}, {de Gregorio-Monsalvo}, {Di
  Francesco}, {Gueth}, {Kawamura}, {Lee}, {Nguyen Luong}, {Mangum}, {Pietu},
  {Sanhueza}, {Saigo}, {Takakuwa}, {Ubach}, {van Kempen}, {Wootten},
  {Castro-Carrizo}, {Francke}, {Gallardo}, {Garcia}, {Gonzalez}, {Hill},
  {Kaminski}, {Kurono}, {Liu}, {Lopez}, {Morales}, {Plarre}, {Schieven},
  {Testi}, {Videla}, {Villard}, {Andreani}, {Hibbard}, \&
  {Tatematsu}}]{ALMA-Partnership2015}
{ALMA Partnership}, {Brogan}, C.~L., {P{\'e}rez}, L.~M., {et~al.} 2015, \apjl,
  808, L3, \dodoi{10.1088/2041-8205/808/1/L3}

\bibitem[{{Andrews} {et~al.}(2016){Andrews}, {Wilner}, {Zhu}, {Birnstiel},
  {Carpenter}, {P{\'e}rez}, {Bai}, {{\"O}berg}, {Hughes}, {Isella}, \&
  {Ricci}}]{Andrews2016}
{Andrews}, S.~M., {Wilner}, D.~J., {Zhu}, Z., {et~al.} 2016, \apjl, 820, L40,
  \dodoi{10.3847/2041-8205/820/2/L40}

\bibitem[{{Ayliffe} \& {Bate}(2009)}]{Ayliffe2009}
{Ayliffe}, B.~A., \& {Bate}, M.~R. 2009, \mnras, 397, 657,
  \dodoi{10.1111/j.1365-2966.2009.15002.x}

\bibitem[{{Ayliffe} \& {Bate}(2010)}]{Ayliffe2010}
---. 2010, \mnras, 408, 876, \dodoi{10.1111/j.1365-2966.2010.17221.x}

\bibitem[{{Bate} {et~al.}(1995){Bate}, {Bonnell}, \& {Price}}]{Bate95}
{Bate}, M.~R., {Bonnell}, I.~A., \& {Price}, N.~M. 1995, \mnras, 277, 362

\bibitem[{{Benisty} {et~al.}(2015){Benisty}, {Juhasz}, {Boccaletti},
  {Avenhaus}, {Milli}, {Thalmann}, {Dominik}, {Pinilla}, {Buenzli}, {Pohl},
  {Beuzit}, {Birnstiel}, {de Boer}, {Bonnefoy}, {Chauvin}, {Christiaens},
  {Garufi}, {Grady}, {Henning}, {Huelamo}, {Isella}, {Langlois}, {M{\'e}nard},
  {Mouillet}, {Olofsson}, {Pantin}, {Pinte}, \& {Pueyo}}]{Benisty2015}
{Benisty}, M., {Juhasz}, A., {Boccaletti}, A., {et~al.} 2015, \aap, 578, L6

\bibitem[{{B{\'e}thune} {et~al.}(2016){B{\'e}thune}, {Lesur}, \&
  {Ferreira}}]{Bethune2016}
{B{\'e}thune}, W., {Lesur}, G., \& {Ferreira}, J. 2016, \aap, 589, A87

\bibitem[{{Biller} {et~al.}(2014){Biller}, {Males}, {Rodigas}, {Morzinski},
  {Close}, {Juh{\'a}sz}, {Follette}, {Lacour}, {Benisty}, {Sicilia-Aguilar},
  {Hinz}, {Weinberger}, {Henning}, {Pott}, {Bonnefoy}, \&
  {K{\"o}hler}}]{Biller2014}
{Biller}, B.~A., {Males}, J., {Rodigas}, T., {et~al.} 2014, \apjl, 792, L22,
  \dodoi{10.1088/2041-8205/792/1/L22}

\bibitem[{{Brittain} {et~al.}(2014){Brittain}, {Carr}, {Najita}, {Quanz}, \&
  {Meyer}}]{Brittain2014}
{Brittain}, S.~D., {Carr}, J.~S., {Najita}, J.~R., {Quanz}, S.~P., \& {Meyer},
  M.~R. 2014, \apj, 791, 136, \dodoi{10.1088/0004-637X/791/2/136}

\bibitem[{{Casassus} {et~al.}(2015){Casassus}, {Marino}, {P{\'e}rez}, {Roman},
  {Dunhill}, {Armitage}, {Cuadra}, {Wootten}, {van der Plas}, {Cieza}, {Moral},
  {Christiaens}, \& {Montesinos}}]{Casassus2015a}
{Casassus}, S., {Marino}, S., {P{\'e}rez}, S., {et~al.} 2015, \apj, 811, 92

\bibitem[{{Currie} {et~al.}(2015){Currie}, {Cloutier}, {Brittain}, {Grady},
  {Burrows}, {Muto}, {Kenyon}, \& {Kuchner}}]{Currie2015}
{Currie}, T., {Cloutier}, R., {Brittain}, S., {et~al.} 2015, \apjl, 814, L27,
  \dodoi{10.1088/2041-8205/814/2/L27}

\bibitem[{{de Gregorio-Monsalvo} {et~al.}(2013){de Gregorio-Monsalvo},
  {M{\'e}nard}, {Dent}, {Pinte}, {L{\'o}pez}, {Klaassen}, {Hales},
  {Cort{\'e}s}, {Rawlings}, {Tachihara}, {Testi}, {Takahashi}, {Chapillon},
  {Mathews}, {Juhasz}, {Akiyama}, {Higuchi}, {Saito}, {Nyman}, {Phillips},
  {Rod{\'o}n}, {Corder}, \& {Van Kempen}}]{de-Gregorio-Monsalvo2013}
{de Gregorio-Monsalvo}, I., {M{\'e}nard}, F., {Dent}, W., {et~al.} 2013, \aap,
  557, A133

\bibitem[{{Dipierro} {et~al.}(2015){Dipierro}, {Pinilla}, {Lodato}, \&
  {Testi}}]{Dipierro2015a}
{Dipierro}, G., {Pinilla}, P., {Lodato}, G., \& {Testi}, L. 2015, \mnras, 451,
  974, \dodoi{10.1093/mnras/stv970}

\bibitem[{{Draine}(2003)}]{Draine03}
{Draine}, B.~T. 2003, \apj, 598, 1017, \dodoi{10.1086/379118}

\bibitem[{{Flaherty} {et~al.}(2015){Flaherty}, {Hughes}, {Rosenfeld},
  {Andrews}, {Chiang}, {Simon}, {Kerzner}, \& {Wilner}}]{Flaherty2015}
{Flaherty}, K.~M., {Hughes}, A.~M., {Rosenfeld}, K.~A., {et~al.} 2015, \apj,
  813, 99

\bibitem[{{Flaherty} {et~al.}(2017){Flaherty}, {Hughes}, {Rose}, {Simon}, {Qi},
  {Andrews}, {K{\'o}sp{\'a}l}, {Wilner}, {Chiang}, {Armitage}, \&
  {Bai}}]{Flaherty2017}
{Flaherty}, K.~M., {Hughes}, A.~M., {Rose}, S.~C., {et~al.} 2017, \apj, 843,
  150

\bibitem[{{Flock} {et~al.}(2015){Flock}, {Ruge}, {Dzyurkevich}, {Henning},
  {Klahr}, \& {Wolf}}]{Flock2015}
{Flock}, M., {Ruge}, J.~P., {Dzyurkevich}, N., {et~al.} 2015, \aap, 574, A68,
  \dodoi{10.1051/0004-6361/201424693}

\bibitem[{{Gaia Collaboration} {et~al.}(2018){Gaia Collaboration}, {Brown},
  {Vallenari}, {Prusti}, {de Bruijne}, {Babusiaux}, \&
  {Bailer-Jones}}]{Gaia-Collaboration2018}
{Gaia Collaboration}, {Brown}, A.~G.~A., {Vallenari}, A., {et~al.} 2018, ArXiv
  e-prints.
\newblock \doarXiv{1804.09365}

\bibitem[{{Gonzalez} {et~al.}(2015){Gonzalez}, {Laibe}, {Maddison}, {Pinte}, \&
  {M{\'e}nard}}]{Gonzalez2015}
{Gonzalez}, J.-F., {Laibe}, G., {Maddison}, S.~T., {Pinte}, C., \&
  {M{\'e}nard}, F. 2015, \mnras, 454, L36, \dodoi{10.1093/mnrasl/slv120}

\bibitem[{{Grady} {et~al.}(2000){Grady}, {Devine}, {Woodgate}, {Kimble},
  {Bruhweiler}, {Boggess}, {Linsky}, {Plait}, {Clampin}, \&
  {Kalas}}]{Grady2000}
{Grady}, C.~A., {Devine}, D., {Woodgate}, B., {et~al.} 2000, \apj, 544, 895

\bibitem[{{Guidi} {et~al.}}(2018){Guidi G. et~al.}]{Guidi2018}
{Guidi, G. et al.} 2018, submitted

\bibitem[{{Isella} {et~al.}(2016){Isella}, {Guidi}, {Testi}, {Liu}, {Li}, {Li},
  {Weaver}, {Boehler}, {Carperter}, {De Gregorio-Monsalvo}, {Manara}, {Natta},
  {P{\'e}rez}, {Ricci}, {Sargent}, {Tazzari}, \& {Turner}}]{Isella2016a}
{Isella}, A., {Guidi}, G., {Testi}, L., {et~al.} 2016, Physical Review Letters,
  117, 251101

\bibitem[{{Kraus} \& {Ireland}(2012)}]{Kraus2012}
{Kraus}, A.~L., \& {Ireland}, M.~J. 2012, \apj, 745, 5,
  \dodoi{10.1088/0004-637X/745/1/5}

\bibitem[{{Ligi} {et~al.}(2018){Ligi}, {Vigan}, {Gratton}, {de Boer},
  {Benisty}, {Boccaletti}, {Quanz}, {Meyer}, {Ginski}, {Sissa}, {Gry},
  {Henning}, {Beuzit}, {Biller}, {Bonnefoy}, {Chauvin}, {Cheetham}, {Cudel},
  {Delorme}, {Desidera}, {Feldt}, {Galicher}, {Girard}, {Janson}, {Kasper},
  {Kopytova}, {Lagrange}, {Langlois}, {Lecoroller}, {Maire}, {M{\'e}nard},
  {Mesa}, {Peretti}, {Perrot}, {Pinilla}, {Pohl}, {Rouan}, {Stolker},
  {Samland}, {Wahhaj}, {Wildi}, {Zurlo}, {Buey}, {Fantinel}, {Fusco}, {Jaquet},
  {Moulin}, {Ramos}, {Suarez}, \& {Weber}}]{Ligi2018}
{Ligi}, R., {Vigan}, A., {Gratton}, R., {et~al.} 2018, \mnras, 473, 1774,
  \dodoi{10.1093/mnras/stx2318}

\bibitem[{{Liu} {et~al.}(2018){Liu}, {Jin}, {Li}, {Isella}, \& {Li}}]{Liu2018}
{Liu}, S.-F., {Jin}, S., {Li}, S., {Isella}, A., \& {Li}, H. 2018, ArXiv
  e-prints.
\newblock \doarXiv{1803.05437}

\bibitem[{{Lodato} \& {Price}(2010)}]{Lodato2010}
{Lodato}, G., \& {Price}, D.~J. 2010, \mnras, 405, 1212

\bibitem[{{Lor{\'e}n-Aguilar} \& {Bate}(2015)}]{Loren2015}
{Lor{\'e}n-Aguilar}, P., \& {Bate}, M.~R. 2015, \mnras, 453, L78,
  \dodoi{10.1093/mnrasl/slv109}

\bibitem[{{McMullin} {et~al.}(2007){McMullin}, {Waters}, {Schiebel}, {Young},
  \& {Golap}}]{McMullin07}
{McMullin}, J.~P., {Waters}, B., {Schiebel}, D., {Young}, W., \& {Golap}, K.
  2007, in Astronomical Society of the Pacific Conference Series, Vol. 376,
  Astronomical Data Analysis Software and Systems XVI, ed. R.~A. {Shaw},
  F.~{Hill}, \& D.~J. {Bell}, 127

\bibitem[{{Muro-Arena} {et~al.}(2018){Muro-Arena}, {Dominik}, {Waters}, {Min},
  {Klarmann}, {Ginski}, {Isella}, {Benisty}, {Pohl}, {Garufi}, {Hagelberg},
  {Langlois}, {Menard}, {Pinte}, {Sezestre}, {van der Plas}, {Villenave},
  {Delboulb{\'e}}, {Magnard}, {M{\"o}ller-Nilsson}, {Pragt}, {Rabou}, \&
  {Roelfsema}}]{Muro-Arena2018}
{Muro-Arena}, G.~A., {Dominik}, C., {Waters}, L.~B.~F.~M., {et~al.} 2018, ArXiv
  e-prints.
\newblock \doarXiv{1802.03328}

\bibitem[{{Natta} {et~al.}(2004){Natta}, {Testi}, {Neri}, {Shepherd}, \&
  {Wilner}}]{Natta2004}
{Natta}, A., {Testi}, L., {Neri}, R., {Shepherd}, D.~S., \& {Wilner}, D.~J.
  2004, \aap, 416, 179, \dodoi{10.1051/0004-6361:20035620}

\bibitem[{{Ogilvie} \& {Lubow}(2002)}]{Ogilvie2002}
{Ogilvie}, G.~I., \& {Lubow}, S.~H. 2002, \mnras, 330, 950,
  \dodoi{10.1046/j.1365-8711.2002.05148.x}

\bibitem[{{Perez} {et~al.}(2015){Perez}, {Dunhill}, {Casassus}, {Roman},
  {Szul{\'a}gyi}, {Flores}, {Marino}, \& {Montesinos}}]{Perez2015b}
{Perez}, S., {Dunhill}, A., {Casassus}, S., {et~al.} 2015, \apjl, 811, L5

\bibitem[{{Pinte} {et~al.}(2009){Pinte}, {Harries}, {Min}, {Watson},
  {Dullemond}, {Woitke}, {M{\'e}nard}, \& {Dur{\'a}n-Rojas}}]{Pinte09}
{Pinte}, C., {Harries}, T.~J., {Min}, M., {et~al.} 2009, \aap, 498, 967,
  \dodoi{10.1051/0004-6361/200811555}

\bibitem[{{Pinte} {et~al.}(2006){Pinte}, {M{\'e}nard}, {Duch{\^e}ne}, \&
  {Bastien}}]{Pinte06}
{Pinte}, C., {M{\'e}nard}, F., {Duch{\^e}ne}, G., \& {Bastien}, P. 2006, \aap,
  459, 797, \dodoi{10.1051/0004-6361:20053275}

\bibitem[{{Pinte} {et~al.}(2018){Pinte}, {M{\'e}nard}, {Duch{\^e}ne}, {Hill},
  {Dent}, {Woitke}, {Maret}, {van der Plas}, {Hales}, {Kamp}, {Thi}, {de
  Gregorio-Monsalvo}, {Rab}, {Quanz}, {Avenhaus}, {Carmona}, \&
  {Casassus}}]{Pinte2018}
{Pinte}, C., {M{\'e}nard}, F., {Duch{\^e}ne}, G., {et~al.} 2018, \aap, 609, A47

\bibitem[{{Price}(2007)}]{Price2007}
{Price}, D.~J. 2007, \pasa, 24, 159

\bibitem[{{Price} {et~al.}(2017){Price}, {Wurster}, {Nixon}, {Tricco},
  {Toupin}, {Pettitt}, {Chan}, {Laibe}, {Glover}, {Dobbs}, {Nealon}, {Liptai},
  {Worpel}, {Bonnerot}, {Dipierro}, {Ragusa}, {Federrath}, {Iaconi},
  {Reichardt}, {Forgan}, {Hutchison}, {Constantino}, {Ayliffe}, {Mentiplay},
  {Hirsh}, \& {Lodato}}]{Price2017}
{Price}, D.~J., {Wurster}, J., {Nixon}, C., {et~al.} 2017, ArXiv e-prints.
\newblock \doarXiv{1702.03930}

\bibitem[{{Price} {et~al.}(2018){Price}, {Cuello}, {Pinte}, {Mentiplay},
  {Casassus}, {Christiaens}, {Kennedy}, {Cuadra}, {Perez}, {Marino},
  {Armitage}, {Zurlo}, {Juhasz}, {Ragusa}, {Laibe}, \& {Lodato}}]{Price2018}
{Price}, D.~J., {Cuello}, N., {Pinte}, C., {et~al.} 2018, \mnras,
  \dodoi{10.1093/mnras/sty647}

\bibitem[{{Quanz} {et~al.}(2015){Quanz}, {Amara}, {Meyer}, {Girard},
  {Kenworthy}, \& {Kasper}}]{Quanz2015}
{Quanz}, S.~P., {Amara}, A., {Meyer}, M.~R., {et~al.} 2015, \apj, 807, 64,
  \dodoi{10.1088/0004-637X/807/1/64}

\bibitem[{{Quanz} {et~al.}(2013{\natexlab{a}}){Quanz}, {Amara}, {Meyer},
  {Kenworthy}, {Kasper}, \& {Girard}}]{Quanz2013a}
---. 2013{\natexlab{a}}, \apjl, 766, L1, \dodoi{10.1088/2041-8205/766/1/L1}

\bibitem[{{Quanz} {et~al.}(2013{\natexlab{b}}){Quanz}, {Avenhaus}, {Buenzli},
  {Garufi}, {Schmid}, \& {Wolf}}]{Quanz2013b}
{Quanz}, S.~P., {Avenhaus}, H., {Buenzli}, E., {et~al.} 2013{\natexlab{b}},
  \apjl, 766, L2, \dodoi{10.1088/2041-8205/766/1/L2}

\bibitem[{{Rameau} {et~al.}(2017){Rameau}, {Follette}, {Pueyo}, {Marois},
  {Macintosh}, {Millar-Blanchaer}, {Wang}, {Vega}, {Doyon}, {Lafreni{\`e}re},
  {Nielsen}, {Bailey}, {Chilcote}, {Close}, {Esposito}, {Males}, {Metchev},
  {Morzinski}, {Ruffio}, {Wolff}, {Ammons}, {Barman}, {Bulger}, {Cotten}, {De
  Rosa}, {Duchene}, {Fitzgerald}, {Goodsell}, {Graham}, {Greenbaum}, {Hibon},
  {Hung}, {Ingraham}, {Kalas}, {Konopacky}, {Larkin}, {Maire}, {Marchis},
  {Oppenheimer}, {Palmer}, {Patience}, {Perrin}, {Poyneer}, {Rajan},
  {Rantakyr{\"o}}, {Marley}, {Savransky}, {Schneider}, {Sivaramakrishnan},
  {Song}, {Soummer}, {Thomas}, {Wallace}, {Ward-Duong}, \&
  {Wiktorowicz}}]{Rameau2017}
{Rameau}, J., {Follette}, K.~B., {Pueyo}, L., {et~al.} 2017, \aj, 153, 244,
  \dodoi{10.3847/1538-3881/aa6cae}

\bibitem[{{Reggiani} {et~al.}(2014){Reggiani}, {Quanz}, {Meyer}, {Pueyo},
  {Absil}, {Amara}, {Anglada}, {Avenhaus}, {Girard}, {Carrasco Gonzalez},
  {Graham}, {Mawet}, {Meru}, {Milli}, {Osorio}, {Wolff}, \&
  {Torrelles}}]{Reggiani2014}
{Reggiani}, M., {Quanz}, S.~P., {Meyer}, M.~R., {et~al.} 2014, \apjl, 792, L23,
  \dodoi{10.1088/2041-8205/792/1/L23}

\bibitem[{{Reggiani} {et~al.}(2018){Reggiani}, {Christiaens}, {Absil}, {Mawet},
  {Huby}, {Choquet}, {Gomez Gonzalez}, {Ruane}, {Femenia}, {Serabyn},
  {Matthews}, {Barraza}, {Carlomagno}, {Defr{\`e}re}, {Delacroix}, {Habraken},
  {Jolivet}, {Karlsson}, {Orban de Xivry}, {Piron}, {Surdej}, {Vargas Catalan},
  \& {Wertz}}]{Reggiani2018}
{Reggiani}, M., {Christiaens}, V., {Absil}, O., {et~al.} 2018, \aap, 611, A74,
  \dodoi{10.1051/0004-6361/201732016}

\bibitem[{{Rosenfeld} {et~al.}(2013){Rosenfeld}, {Andrews}, {Hughes}, {Wilner},
  \& {Qi}}]{Rosenfeld2013a}
{Rosenfeld}, K.~A., {Andrews}, S.~M., {Hughes}, A.~M., {Wilner}, D.~J., \&
  {Qi}, C. 2013, \apj, 774, 16

\bibitem[{{Sallum} {et~al.}(2015){Sallum}, {Follette}, {Eisner}, {Close},
  {Hinz}, {Kratter}, {Males}, {Skemer}, {Macintosh}, {Tuthill}, {Bailey},
  {Defr{\`e}re}, {Morzinski}, {Rodigas}, {Spalding}, {Vaz}, \&
  {Weinberger}}]{Sallum2015}
{Sallum}, S., {Follette}, K.~B., {Eisner}, J.~A., {et~al.} 2015, \nat, 527,
  342, \dodoi{10.1038/nature15761}

\bibitem[{{Shakura} \& {Sunyaev}(1973)}]{ShakuraSunyaev1973}
{Shakura}, N.~I., \& {Sunyaev}, R.~A. 1973, \aap, 24, 337

\bibitem[{{Stolker} {et~al.}(2016){Stolker}, {Dominik}, {Avenhaus}, {Min}, {de
  Boer}, {Ginski}, {Schmid}, {Juhasz}, {Bazzon}, {Waters}, {Garufi},
  {Augereau}, {Benisty}, {Boccaletti}, {Henning}, {Langlois}, {Maire},
  {M{\'e}nard}, {Meyer}, {Pinte}, {Quanz}, {Thalmann}, {Beuzit}, {Carbillet},
  {Costille}, {Dohlen}, {Feldt}, {Gisler}, {Mouillet}, {Pavlov}, {Perret},
  {Petit}, {Pragt}, {Rochat}, {Roelfsema}, {Salasnich}, {Soenke}, \&
  {Wildi}}]{Stolker2016}
{Stolker}, T., {Dominik}, C., {Avenhaus}, H., {et~al.} 2016, ArXiv e-prints

\bibitem[{{Takahashi} \& {Inutsuka}(2014)}]{Takahashi2014}
{Takahashi}, S.~Z., \& {Inutsuka}, S.-i. 2014, \apj, 794, 55,
  \dodoi{10.1088/0004-637X/794/1/55}

\bibitem[{{Thalmann} {et~al.}(2015){Thalmann}, {Mulders}, {Janson}, {Olofsson},
  {Benisty}, {Avenhaus}, {Quanz}, {Schmid}, {Henning}, {Buenzli}, {M{\'e}nard},
  {Carson}, {Garufi}, {Messina}, {Dominik}, {Leisenring}, {Chauvin}, \&
  {Meyer}}]{Thalmann2015}
{Thalmann}, C., {Mulders}, G.~D., {Janson}, M., {et~al.} 2015, \apjl, 808, L41,
  \dodoi{10.1088/2041-8205/808/2/L41}

\bibitem[{{Thalmann} {et~al.}(2016){Thalmann}, {Janson}, {Garufi},
  {Boccaletti}, {Quanz}, {Sissa}, {Gratton}, {Salter}, {Benisty}, {Bonnefoy},
  {Chauvin}, {Daemgen}, {Desidera}, {Dominik}, {Engler}, {Feldt}, {Henning},
  {Lagrange}, {Langlois}, {Lannier}, {Le Coroller}, {Ligi}, {M{\'e}nard},
  {Mesa}, {Meyer}, {Mulders}, {Olofsson}, {Pinte}, {Schmid}, {Vigan}, \&
  {Zurlo}}]{Thalmann2016}
{Thalmann}, C., {Janson}, M., {Garufi}, A., {et~al.} 2016, \apjl, 828, L17,
  \dodoi{10.3847/2041-8205/828/2/L17}

\bibitem[{{Walsh} {et~al.}(2017){Walsh}, {Daley}, {Facchini}, \&
  {Juh{\'a}sz}}]{Walsh2017}
{Walsh}, C., {Daley}, C., {Facchini}, S., \& {Juh{\'a}sz}, A. 2017, \aap, 607,
  A114, \dodoi{10.1051/0004-6361/201731334}

\bibitem[{{Weaver} {et~al.}(2018){Weaver}, {Isella}, \& {Boehler}}]{Weaver2018}
{Weaver}, E., {Isella}, A., \& {Boehler}, Y. 2018, \apj, 853, 113,
  \dodoi{10.3847/1538-4357/aaa481}

\bibitem[{{Zhang} {et~al.}(2015){Zhang}, {Blake}, \& {Bergin}}]{Zhang2015}
{Zhang}, K., {Blake}, G.~A., \& {Bergin}, E.~A. 2015, \apjl, 806, L7,
  \dodoi{10.1088/2041-8205/806/1/L7}

\end{thebibliography}

\listofchanges

\end{document}